\documentclass[
  10pt, twocolumn,
]{article}

\title {\LARGE\bf   Core percolation in random graphs: \\ 
                    a critical phenomena analysis}

\author {           M. Bauer and O. Golinelli,
\bigskip
\\ \ad              email: \{bauer, golinelli\}@spht.saclay.cea.fr
\\ \ad              Service de Physique Th\'eorique, Cea Saclay,
                    91191 Gif-sur-Yvette, France
}
\date{\normalsize   January 31, 2001; revised June 22, 2001.
\\                  Preprint T01/017; arXiv:cond-mat/0102011.
\bigskip
\\                  Eur. Phys. J. B {\bf 24}, 339--352 (2001).
}

\newcommand  {\ad}{\normalsize\em}      
\usepackage[final]{graphicx}
\newcommand{\figwidth}{\columnwidth}       

\newcommand{\bin}[2]{{ #1 \choose #2 }} 
\newcommand{\Var}{\mbox{Var}}

\begin{document}

\twocolumn[\hsize\textwidth\columnwidth\hsize\csname @twocolumnfalse\endcsname
\maketitle
\begin{abstract} \normalsize
We study both numerically and analytically what happens to a random graph
of average connectivity $\alpha$ when its leaves and their neighbors are
removed iteratively up to the point when no leaf remains.  The remnant is
made of isolated vertices plus an induced subgraph we call \emph{the core}.
In the thermodynamic limit of an infinite random graph, we compute
analytically the dynamics of leaf removal, the number of isolated vertices
and the number of vertices and edges in the core.  We show that a second
order phase transition occurs at $\alpha = e = 2.718\dots$: below the
transition, the core is small but above the transition, it occupies a
finite fraction of the initial graph.  The finite size scaling properties
are then studied numerically in detail in the critical region, and we
propose a consistent set of critical exponents, which does not coincide
with the set of standard percolation exponents for this model.  We clarify
several aspects in combinatorial optimization and spectral properties of
the adjacency matrix of random graphs.


\end{abstract}
\vskip2pc ]              

\section{Introduction}

What remains of a graph when leaves are iteratively removed until none 
remains ? The answer depends on what is meant by leaves. 

In the most standard definition, a leaf is a vertex with exactly one
neighbor, and leaf removal deletes this vertex and the adjacent edge.
In the context of large random graphs where the connectivity $\alpha$
(the average number of neighbors of a vertex) is kept fixed and the
number of vertices $N \to \infty$, the answer is well known and
interesting.  When $\alpha < 1$, the remnant after leaf removal is
made of $O(N)$ isolated vertices, plus a subgraph of size $o(N)$
without leaves. When $\alpha >1$, the remnant still contains $O(N)$
isolated points, but the rest is a subgraph of size $O(N)$, which is
dominated by a single connected component usually called the backbone.
 The a priori surprising, but rather general, fact that backbone
percolation and standard percolation occur at the same point, namely
at $\alpha=1$, has a very simple explanation for random graphs. 
Indeed, a large random graph of average connectivity $\alpha<1$
consists of a forest (union of finite trees) plus a finite number of
finite connected components with one closed loop. Obviously, each tree
shrinks to a single isolated point after leaf removal. However, at
$\alpha = 1$ the percolation transition occurs and when $\alpha >1$, a
random graph consists of a forest plus a finite number of components
with one closed loop, plus a ``giant'' connected component containing
a finite fraction of the vertices and an extensive number of loops. No
loop is destroyed by leaf removal so that the giant component leads to
a macroscopic connected remnant after leaf removal.  The
percolation transition was discovered and studied by Erd\"os and R\'enyi
in their seminal paper~\cite{erdos}. This has initiated a lot of work
on the random graph model, and many fine details concerning the
structure of the percolation transition have been computed (see e.g
Ref.~\cite{knuth}).

The random graph model is believed to be essentially equivalent to
a mean field approximation for percolation on (finite dimensional)
lattices, leading to critical exponents which are valid above the
upper critical dimension, which is $d_c=6$ for percolation.

In this paper, we consider the removal of a slightly more complicated
pattern: we remove at each step not only the leaf but also its
neighbor (and consequently all adjacent edges).  To avoid cumbersome
circumlocutions, in the rest of this paper, we call \emph{leaf} the
pair ``standard leaf + its neighbor''.  Now leaf removal deletes two
vertices (a vertex with a single neighbor and this neighbor) and all
the edges adjacent to one or both vertices.  It is quite natural to
study the removal of these patterns because it has a number of
applications to graph theory: several numerical characteristics of a
graph behave nicely under leaf removal. One such characteristic, which
was our original motivation from physics, is the multiplicity of the
eigenvalue $0$ in the adjacency matrix of the graph. Others are the
minimal size of a vertex cover and the maximal size of an edge
disjoint subset (the matching problem), questions which
are related to various combinatorial optimization problems.

The matching problem had already led mathematicians to a thorough
study of leaf removal (see Refs.~\cite{karpsipser,pitteletal} and
references therein). In fact, parts of our analytical results have
already been obtained in this context. However, we have obtained them
independently by a direct enumeration technique which turned out to be
quite similar to a counting lemma for bicolored trees that appeared
in \cite{bollobright}.

The main result on the structure of the remnant after iterated leaf
removal when the graph is a large random graph of finite connectivity
$\alpha$ is the following. The residue consists of $i(\alpha)N+O(1)$
isolated points and an induced subgraph without leaves or isolated
points which we call the core. It contains $c(\alpha)N+O(1)$ vertices
and $l(\alpha)N+O(1)$ edges. For $\alpha < e = 2.718\dots$,
$c(\alpha)=l(\alpha)=0$ so the core is small. A second order phase
transition occurs at $\alpha=e$ and for $\alpha >e$, $c(\alpha)$ and
$l(\alpha)$ are $>0$. We shall argue that the core is made of a giant
connected ``core'' component plus a finite number of small connected
components involving a total of $o(N)$ vertices. The function
$i(\alpha)$ is always non-vanishing, but it is non-analytic at
$\alpha=e$.

The phase transition at $\alpha=e$ was found initially for the
matching problem \cite{karpsipser,pitteletal}. Physicists however have
observed independently that some properties of random graphs are
singular at $\alpha=e$ (see Ref.~\cite{hartweig1} for replica symmetry
breaking in minimal vertex covers, Ref.~\cite{baugolloctrans} for a
localization problem, and, in a numerical context,
Ref.~\cite{baugolmom} where an anomaly close to the eigenvalue $0$ in
the spectrum of random adjacency matrices was observed). 
 
\vspace{.3cm}

The paper is organized as follows.  The general definitions have been
regrouped in section \ref{sec:defs}.  They are standard and should be used
only for reference.

In section \ref{sec:core} we define leaf removal, leaf removal processes
obtained by iteration of leaf removals and the ``core'' for an arbitrary
graph.

Section \ref{sec:coreperco} presents our derivation of the main results for
large random graphs.  The analytical formul\ae\ for $i(\alpha)$,
$c(\alpha)$ and $l(\alpha)$ are given.

In section \ref{sec:numsim}, these formul\ae\ are checked against
Monte-Carlo simulations of leaf removal processes, which we also use for
the finite size scaling analysis in the critical region.  This leads us to
the definition and numerical evaluation of many new critical exponents. In
particular, we give good evidence that the core percolation exponents (at
$\alpha =e$) are not the same as the critical exponents of standard
percolation (at $\alpha =1$). Even if core percolation on a random graph
can presumably be seen as a mean field approximation for core percolation
on (finite dimensional) lattices with impurities, the corresponding
effective field theory and its upper critical dimension are not known to
us.

In section \ref{sec:applis}, we give two applications of our results.  We
show in particular in section \ref{sec:loc} that for any $\alpha$ the core
of a random graph only carries a small number of 0 eigenvalues of the
adjacency matrix and that the emergence of the core has a direct impact on
the localized and delocalized eigenvectors with eigenvalue 0.  In section
\ref{sec:guard}, we show that for $\alpha < e$, the problem of finding
minimal vertex covers or maximal edge disjoint subsets (matchings) can be
handled very simply in polynomial time (in fact, in linear time once the
graph is encoded in a suitable form).  While the matching problem can
always be solved in polynomial time, the minimal vertex cover problem is
believed to be NP-hard for general graphs, and the same should be true on
the core of a random graph for $\alpha > e$.

The formal proof that the core is a well-defined object is given in
appendix.

\section{General definitions}
\label{sec:defs}

We start with a few standard definitions.  This section should be used only
for reference.

\textbf{Graph}: A graph (also called a \emph{simple undirected graph} in the
mathematical literature) $G$ is a pair consisting of a set $V$ called the
set of vertices of $G$ and a set $E$ called the set of edges of $G$, whose
elements are pairs of distinct elements of $V$. If $\{v,w\}$ is an edge,
the vertices $v$ and $w$ are called \emph{adjacent} or
\emph{neighbors}. They are the extremities of the edge $\{v,w\}$.  Note
that there is at most one edge between two vertices, and that there is no
edge connecting a vertex with itself: the word \emph{simple} above refers
to these two restrictions.

\textbf{Adjacency matrix}: The adjacency matrix of a graph $G$ is a square
matrix $M_{v,w}$ indexed by vertices of $G$ and such that $M_{v,w}=1$ if
$\{v,w\}$ is an edge of $G$ and $0$ otherwise. Note that $M$ is a symmetric
$0-1$ matrix with zeroes on the diagonal. Conversely, any such matrix is
the adjacency matrix of a graph. We denote by $Z(G)$ the dimension of the
kernel (that is, the subspace of eigenvectors with eigenvalue $0$) of the
adjacency matrix of $G$.

\textbf{Induced subgraph}: If $V' \subset V$, the graph with vertex set
$V'$ and edge set $E'$ those edges in $E$ with both extremities in $V'$ is
called the subgraph of $G$ induced by $V'$.

\textbf{Random graph in the microcanonical ensemble}: If
$V=\{1,\cdots,N\}$, there are $\bin{N(N-1)/2}{L}$ graphs with vertex set
$V$ and $L$ edges (making a total of $2^{N(N-1)/2}$ graphs with vertex set
$V$). Saying that all $\bin{N(N-1)/2}{L}$ are equiprobable turns the set of
graphs on $N$ vertices and $L$ edges into a probability space whose
elements we call \emph{random graphs in the microcanonical ensemble}. This
is the ensemble we use below for numerical simulations.

\textbf{Random graph in the canonical ensemble}: Given a number $p \in
[0,1]$, we introduce $\frac{N(N-1)}{2}$ independent random variables
$\varepsilon_{i,j}$, $1\leq i < j \leq N$, each taking value $1$ with
probability $p$ and $0$ with probability $1-p$. Saying that $\{i,j\}$ is an
edge of $G$ if and only if $\varepsilon_{i,j}=1$ turns the set of all
$2^{N(N-1)/2}$ graphs with vertex set $V$ into a probability space whose
elements we call \emph{random graphs in the canonical ensemble}. This is
the ensemble we use below for analytical computations.

\textbf{Connectivity, $\alpha$}: In the sequel, we are interested in the
large $N$ limit with a finite limit $\alpha$ for $\frac{2L}{N}$
(microcanonical ensemble) or for $p(N-1)$ (canonical ensemble). The
parameter $\alpha$ is the average connectivity, the average number of
neighbors of a given vertex in the random graph.

If $N \left( p(1-p) \right)^{1/2} \to \infty$ and $L \sim
p\frac{N(N-1)}{2}$, the thermodynamic properties of a $G(N,L)$ in the
microcanonical ensemble and of a $G(N,p)$ in the canonical ensemble are the
same.  This is in particular true if $pN = \alpha$ is kept fixed as $N \to
\infty$.

\section{Leaf removal process and the core of a graph}
\label{sec:core}

Our aim is to define, for any (random or not) finite graph $G$, a
remarkable subgraph which we call the \emph{core} of $G$. It is
obtained by \emph{leaf removal}, an operation that we define now.

\textbf{Leaf}: A leaf of a graph $G$ is a couple of vertices $(v,w)$ such
that $\{v,w\}$ is an edge of $G$ and $w$ belongs to no other edge of $G$.
Note that this is not the most standard definition and that $(v,w)$ and
$(w,v)$ are both leaves if and only if $\{v,w\}$ is a connected component
of $G$.

\textbf{Bunch of leaves}: A bunch of leaves is a maximal family of leaves
with the same first vertex. The leaves of a graph can be grouped into
bunches of leaves in an unique way.

\begin{figure}
  \centering
  \includegraphics[width=\figwidth, height=0.9\textheight, keepaspectratio]
                  {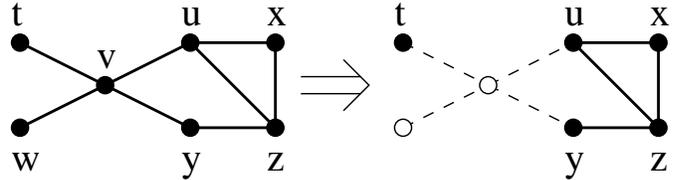}
  \caption{\em In this example, the leaf $(v,w)$ is removed, as well as the
    four edges touching $v$: the new graph $G'$ is the subgraph of $G$
    induced by the five remaining vertices. Note that the vertex $t$ is now
    isolated, and that a new leaf $(z,y)$ has been created.}
  \label{fig:eff}
\end{figure}

\textbf{Leaf removal}: If $(v,w)$ is a leaf of $G$, and $G'$ the subgraph
of $G$ induced by $V \backslash \{v,w\}$, we say that $G'$ is obtained from
$G$ by leaf removal of $(v,w)$. In other words, $G'$ is obtained from $G$
by removing vertices $v$ and $w$, the edge $\{v,w\}$ and all other edges
touching $v$.  Note that this operation can destroy other leaves of $G$ and
also create new leaves.  See Fig.~\ref{fig:eff} for a pictorial example.

\textbf{Step by step leaf removal process}: Start from a graph $G$. If $G$
has no leaves, stop. Else, choose a leaf $(v,w)$ and remove it, leading to
a graph $G'$. If $G'$ has no leaves, stop. Else, choose a leaf $(v',w')$
and remove it. This operation is iterated until no leaf remains.

\textbf{History}: A sequence $G,(v,w),G',(v',w'),\cdots$ associated to a
step by step leaf removal process is called an history.

\textbf{Isolated points, $I$; Core of a graph, $C$}: The last term in
an history starting from $G$ is a graph which splits into a collection
of isolated points $I$, and an induced subgraph $C$ of $G$ without
leaves or isolated points which we call the core of $G$. We denote the
number of points in the core by $N_c$ and the number of edges in the
core by $L_c$.   

\vspace{.2cm}

For these definitions to make sense, one has to show that the
number of isolated points and the core are well defined, that is, do
not depend on the choice of history.  The formal proof is given in
the appendix. 

\vspace{.2cm}

\textbf{Global leaf removal process}: Start from a graph $G$. If it has no
leaves, stop. Else select one leaf in every bunch of leaves.  Remove from
the vertex set $V$ both extremities of all the selected leaves, and define
$V^{(1)}$ to be the set of remaining vertices. Let $G^{(1)}$ be the
subgraph of $G$ induced by $V^{(1)}$.  Note that the leaves of $G^{(1)}$
(if any) are not leaves of $G$.  In this operation, the vertices belonging
to a bunch of $G$ that were not selected become isolated points of
$G^{(1)}$.  They remain isolated for the rest of the process.  Iterate the
procedure and define $G^{(2)},G^{(3)},\cdots$ until a graph without leaves
is obtained.

In the proof given in the appendix that the core is well-defined,
the argument in step $\mathbf{3c}$ implies in particular
that leaf removals that take
place in distinct bunches of a graph commute. It implies that the
global leaf removal process leads to same core and number of isolated
points as any step by step leaf removal process.

While the global leaf removal process is convenient for analytical
computations, the step by step leaf removal process is easier to implement
on the computer.

\section{Core percolation: infinite $N$ results}
\label{sec:coreperco}

The global leaf removal process allows to compute the most salient
characteristics of the leaf removal process, the functions
$i(\alpha)$, $c(\alpha)$ and $l(\alpha)$. Remember that by definition,
the number of isolated points after leaf removal is $Ni(\alpha)+o(N)$,
the number of vertices in the core is $N_c=Nc(\alpha)+o(N)$, and the
number of edges in the core is $L_c=Nl(\alpha) +o(N)$. The clue is an
enumeration of all the configurations on the random graph that
contribute extensively to the fundamental events in the global leaf
removal process at step $n$ (which goes from $G^{(n-1)}$ to
$G^{(n)}$): emergence of a new isolated point, removal of a point,
removal of an edge\footnote{
  The method of Refs.~\cite{karpsipser,pitteletal} relies on approximate
  differential equations that apply to a slightly different model of random
  graphs.  It is very powerful, but less intuitive than the direct
  enumeration method that follows.
}.  This enumeration is
possible because the finite configurations of vertices and edges in the
random graph with extensive multiplicity are tree-like. This implies that
the problem has a recursive structure.  The weight of a tree is chosen so
as to reproduce the correct random graph weight: a vertex has weight
$e^{-\alpha}$ and an edge has weight $\alpha$.  One has to be rather
careful to avoid double counting and omissions; the examination of all
cases is very tedious so we omit the details and simply outline the
strategy.

The key ingredient is a study of leaf removal on
rooted trees. Starting from a rooted tree, we apply the global leaf
removal process, with the convention that even if the root has only
one neighbor, it is not counted as a standard leaf\footnote{
  However, a configuration where a neighbor of the root is a standard leaf
  is treated as usual.
}. We let $p_n,\; n\geq 0$ be the generating function for rooted trees
whose root becomes isolated exactly at step $n$ of the global leaf removal
process, and $q_n,\; n\geq 1$ be the generating function for rooted trees
whose root is removed exactly at step $n$ of the global leaf removal
process. For instance, $p_0$ counts rooted trees with an isolated root,
hence trees with a single vertex, and $p_0=e^{-\alpha}$. As another
example, $q_1$ counts rooted trees whose root touches at least one standard
leaf. Consider the trees whose root touches exactly $k\geq 1$ standard
leaves, and $l\geq 0$ other vertices. These $l$ vertices can be seen as
roots of non-trivial subtrees of the original tree, so by definition they
contribute to $1-p_0$.  So the weight is
\[ e^{-\alpha} \frac{(\alpha e^{-\alpha})^k}{k!}\frac{(\alpha
  (1-p_0))^l}{l!}.\]  
Hence 
\[ q_1=e^{-\alpha} \sum_{k \geq 1} \sum_{l \geq 0} 
\frac{(\alpha e^{-\alpha})^k}{k!}\frac{(\alpha
  (1-p_0))^l}{l!}=1-e^{-\alpha e^{-\alpha}}.\]

In the same way, contributions to $p_n$ or $q_n$ for larger $n'$s can
be analyzed in terms of the trees attached to the neighbors of the
root, and the structure of these trees involves lower order
contributions. In contributions to $p_n$, the root has at least one
neighbor whose attached tree contributes to $q_n$ and any number of
neighbors contributing to $q_1$ or $q_2$ or $\cdots$ or $q_{n-1}$. So
\[p_n=e^{-\alpha}(e^{\alpha q_n}-1)e^{\alpha (q_{n-1}+\cdots+q_1)}.\]
Analogously, in contributions to $q_n$, the root has at least one
neighbor whose attached tree contributes to $p_{n-1}$ and any number of
neighbors none of which contributing to $p_0$ or $p_1$ or $\cdots$ or
$p_{n-1}$. So  
\[q_n=e^{-\alpha}(e^{\alpha p_{n-1}}-1)e^{\alpha
  (1-p_{n-1}-\cdots-p_0)}.\]
These two relations allow it to be shown that
\begin{eqnarray*} p_n & = & e_{2n+1}-e_{2n-1} \quad \mathrm{for} \; n
\geq 0 \\  q_n & = & e_{2n-2}-e_{2n} \quad \mathrm{for} \;
 n \geq 1, \end{eqnarray*}  
where $e_n(\alpha)$ is the sequence of iterated exponentials, defined by 
\begin{equation}
  e_{-1}=0 \quad \mathrm{and} \quad e_n=e^{-\alpha e_{n-1}} \quad 
 \mathrm{for} \; n \geq 0.
 \label{eq:en}
\end{equation}

The events on the random graph that at step $n$ of the global leaf
removal process a given vertex becomes isolated, or that a given
vertex disappears, or that a given edge disappears can all be
interpreted in terms of the previous configurations. In each case, the
different possible contributions have to be taken into account, and
also the rule that a root with a single neighbor can be touched by
leaf removal has to be enforced. We omit this painful case by case
analysis and only state the results.  

The explicit formul\ae\ for extensive contribution to the average
number $Ni_n(\alpha)$ of isolated vertices, $Nc_n(\alpha)$ of non
isolated vertices and $Nl_n(\alpha)$ of edges in $G^{(n)}$ are
\begin{eqnarray} 
  \label{eq:incndn}
  i_n(\alpha) & = & e_{2n+1}+e_{2n}+\alpha e_{2n}e_{2n-1}-1, \nonumber \\
  c_n(\alpha) & = & e_{2n}-e_{2n+1}-\alpha e_{2n}e_{2n-1}+\alpha
  e_{2n-1}^2,  \\ 
  l_n(\alpha) & = & \frac{\alpha}{2}(e_{2n}-e_{2n-1})^2. \nonumber 
\end{eqnarray}

\begin{figure}
  \centering
  \includegraphics[width=\figwidth, height=0.9\textheight, keepaspectratio]
                  {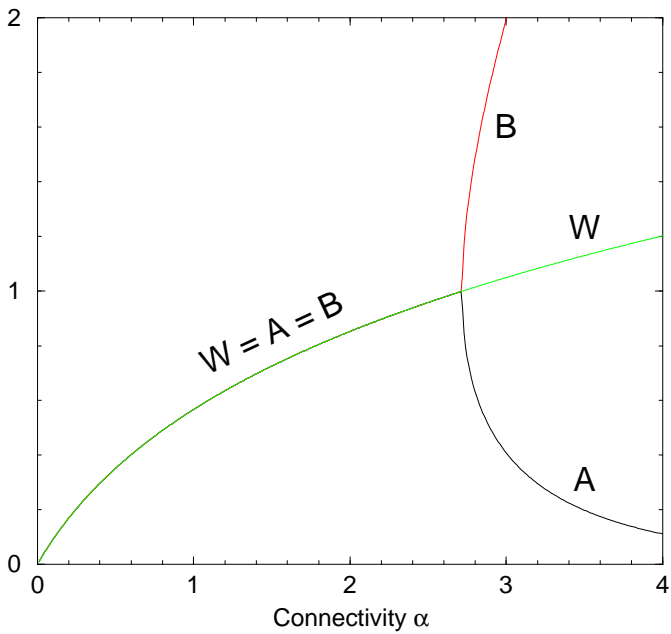}
  \caption{\em 
    The special functions $W(\alpha)$, $A(\alpha)$ and $B(\alpha)$, which
    coincide for $\alpha \leq e$.
  }
  \label{fig:abw}
\end{figure}

Now comes the crucial fact: when $\alpha \leq e$, the sequence
$e_n(\alpha)$ converges to $W(\alpha)/\alpha$, where $W(\alpha)$ is the
Lambert function, defined for $\alpha \geq 0$ as the unique real solution
of the equation $W e^W = \alpha$.  The function $W(\alpha)$ is analytic for
$\alpha \geq 0$.  When $\alpha > e$, $W(\alpha)/\alpha$ remains a fixed
point of the iteration Eq.~(\ref{eq:en}) but it is unstable: the sequence
$\{e_n\}$ is oscillating.  However the even subsequence $e_{2n}$ and odd
subsequence $e_{2n+1}$ are still convergent. The even limit is strictly
larger than the odd limit. We define the functions $A(\alpha)$ and
$B(\alpha)$ for $\alpha \geq 0$ by
\[
  \lim_{n \rightarrow \infty} \, e_{2n} =  \frac{B}{\alpha} \quad
  \mathrm{and} \quad \lim_{n \rightarrow \infty} \, e_{2n+1} = 
  \frac{A}{\alpha}. 
\]
Then $(A,B)$ solve the system 
\begin{equation}
  Ae^B=\alpha, \quad  Be^A=\alpha. 
  \label{eq:ab}
\end{equation}
For $\alpha \leq e$, the unique solution is $A = B = W$.  For $\alpha > e$,
the previous solution becomes unstable and $(A,B)$ is the solution of
Eq.~(\ref{eq:ab}) selected by the rule $A < W < B$.  This is summarized on
Fig.~\ref{fig:abw}.

Taking the limit in Eq.~(\ref{eq:incndn}) leads to 
\begin{eqnarray}
  i(\alpha) & = & \frac{A+B+AB}{\alpha}-1, \nonumber \\
  c(\alpha) & = & \frac{(B-A)(1-A)}{\alpha},  \\ 
  l(\alpha) & = & \frac{(B-A)^2}{2\alpha}.\nonumber 
\end{eqnarray}
For $\alpha \leq e$, $c(\alpha) = l(\alpha) = 0$, and the core indeed
has a size $o(N)$.  On the other hand, the core occupies a finite
fraction $c(\alpha)$ of the vertices for $\alpha >e$.  The behavior of
Eq~(\ref{eq:en}) is responsible for this geometric transition, core
percolation, at $\alpha = e$.  As $c(\alpha)$ and $l(\alpha) \to 0$
when $\alpha \to e^+$, these functions are continuous but their
derivatives are not: the transition is of second order. Note again
that core percolation appears at $\alpha=e$, contrary to backbone
percolation, which occurs at $\alpha =1$.

The fact that the core of a graph is an induced subgraph of the original
graph allows to give a physicist's argument for the uniqueness of the giant
component in the core.  Fix $\alpha >e$. If the core of the random graph
contains two or more large connected components, there was no edge with
extremities in two components in the original graph. But as the total size
of the large connected components is of order $N$, the absence of such an
edge is extremely unlikely.

The behavior of thermodynamic functions close to the transition is
the following. Writing $\alpha=e(1+\varepsilon)$, for small negative
$\varepsilon$,
\begin{eqnarray*}
 \lefteqn{ A(\alpha) = B(\alpha) =}  \\
 && 1+\frac{1}{2}\varepsilon-\frac{3}{16}\varepsilon^2  
           +\frac{19}{192}\varepsilon^3-\frac{185}{3072}\varepsilon^4
           +\frac{2437}{61440}\varepsilon^5+O(\varepsilon^6)
\end{eqnarray*}
while for small positive $\varepsilon$,
\begin{eqnarray*}
 A(\alpha) &\!=\!& 1-6^{1/2}\varepsilon ^{1/2}+2\varepsilon-\frac{6^{1/2}}{20}
 \varepsilon ^{3/2}-\frac{3}{5}\varepsilon ^2 +O(\varepsilon^{3/2}), \\
 B(\alpha) &\!=\!& 1+6^{1/2}\varepsilon ^{1/2}+2\varepsilon+\frac{6^{1/2}}{20}
 \varepsilon ^{3/2}-\frac{3}{5}\varepsilon ^2 +O(\varepsilon^{3/2}).
\end{eqnarray*}
For $i(\alpha)$, this implies that there is a jump only in the second
derivative of at the transition, with
\[ 
 i(\alpha) = \frac{3-e}{e}-\frac{1}{e}\varepsilon+
 \left\{ \begin{array}{ll} 
   \frac{1}{2e}\varepsilon^2+O(\varepsilon^3) & 
   \mathrm{for} \; \varepsilon < 0 \\ 
   & \\
   \frac{2}{e}\varepsilon^2+O(\varepsilon^3) & 
   \mathrm{for} \; \varepsilon > 0 
 \end{array} \right.
\]
while $c(\alpha)$ and $l(\alpha)$ have a jump in the first derivative
at the transition, with
\[ 
 c(\alpha)=\left\{ \begin{array}{ll} 0 & \mathrm{for} \; \varepsilon < 0 \\
 \frac{12}{e}\varepsilon-\frac{4(6)^{1/2}}{e}\varepsilon
 ^{3/2}-\frac{54}{5e}\varepsilon^2 +O(\varepsilon^{5/2}) & \mathrm{for} \;
 \varepsilon > 0 \end{array} 
 \right.
\]
and 
\[ 
 l(\alpha)=\left\{ \begin{array}{ll} 0 & \mathrm{for} \; \varepsilon < 0 \\
 \frac{12}{e}\varepsilon-\frac{54}{5e}\varepsilon^2 +O(\varepsilon ^3)
 & \mathrm{for} \; \varepsilon > 0 . \end{array} 
 \right.
\]
The expansion for $l$ contains only integral powers of $\varepsilon$, and
this may be related to the fact (see section~\ref{sec:numsim}) that the
finite size corrections for the number of edges in the core $L_c$ are
slightly nicer than the ones for the number of vertices in the core
$N_c$. The average connectivity of the core is
\[ 
 \alpha_{eff} = \frac{2l(\alpha)}{c(\alpha)} = \frac{B-A}{1-A}=2+
 \frac{2(6)^{1/2}}{3}\varepsilon ^{1/2}+\frac{4}{3}\varepsilon
 +O(\varepsilon^{3 /2})
\]
for $\varepsilon > 0$, which implies that giant core component should look
like a large loop for $\alpha$ close to $e^+$. The exponent $1/2$ in the
first correction makes such a property quite difficult to see numerically
at large but finite $N$.

In the random graph model, the vertices do not live in any ambient space,
and the notion of correlation length is ambiguous.  This problem will
reappear in the finite size scaling analysis of the next section. However,
the emergence of the core is very reminiscent of critical phenomena in
physics. In particular, the critical slowing down is observable during the
global leaf removal process. Indeed, the speed of convergence of the
iterated exponential sequence can be computed. One finds that for $\alpha
\neq e$, the convergence is exponential: the convergence rate
$\xi^{-1}(\alpha)$ is given  by the formula
\[
  \xi^{-1} =\frac{A+B}{2}-\log \alpha,
\]
and more precisely
\begin{eqnarray*}
 e_{n}-\frac{W}{\alpha} & \sim & (-)^n w e^{-n/\xi} \quad
 \mathrm{for} \; \alpha <e, \\ 
 e_{2n+1}-\frac{A}{\alpha} & \sim & - a e^{-(2n+1)/\xi} \quad
 \mathrm{for} \; \alpha >e, \\
 e_{2n}-\frac{B}{\alpha} & \sim & b e^{-2n/\xi} \quad
 \mathrm{for} \; \alpha >e,
\end{eqnarray*}
where $a(\alpha)$, $w(\alpha)$ and $b(\alpha)$ are positive functions (they
coincide for $\alpha < e$ and $be^{-B/2}=ae^{-A/2}$).  When $\alpha \to
e^-$, $\xi \sim \frac{2e}{e-\alpha}$, and when $\alpha \to e^+$, $\xi \sim
\frac{e}{\alpha-e}$.

At $\alpha=e$, the convergence is algebraic, and
\begin{eqnarray*}
 \lefteqn{ e \left(e_n-\frac{1}{e}\right) = } \\
 & & (-)^n\frac{6^{1/2}}{n^{1/2}}+\frac{3}{2n}+(-)^{n+1} 
     \frac{21(6)^{1/2}}{80} \frac{\log n}{n^{3/2}} 
     +O(\frac{1}{n^{3/2}}).
\end{eqnarray*}

This leads to the asymptotics at $\alpha=e$:
\begin{eqnarray*}
 i_n & = & \frac{3-e}{e}+O\left(\frac{1}{n^{3/2}}\right) , \\
 c_n & = & \frac{6}{en}\left(1-\frac{3^{1/2}}{4n^{1/2}}-\frac{21}{80}
 \frac{\log n}{n} +O\left(\frac{1}{n}\right)\right) , \\
 l_n & = & \frac{6}{en}\left(1-\frac{21}{80}\frac{\log n}{n}
 +O\left(\frac{1}{n}\right)\right).
\end{eqnarray*}
The first correction for $c_n$ is more important that the one for $l_n$.
Moreover the logarithms at $\alpha=e$ lead to suspect that the finite size
analysis of the next section might also be complicated by logarithms.

\section{Numerical studies of the core percolation}
\label{sec:numsim}

The analytical computations above have enabled us to locate a phase
transition at $\alpha = e$. They give information concerning the
critical region but do not exhaust all the critical exponents. So we
made an extensive numerical analysis of the core using Monte-Carlo
simulations.  At the first step this can also be used to check the
previous analytical results.  But let us start with the numerical
algorithm.

\subsection{Monte-Carlo algorithm} 

Our Monte-Carlo simulations consist in generating lots of random graphs,
removing leaves step by step, and studying the remaining cores and isolated
points.  More precisely, for a given set of parameters $(N,\alpha)$, we
generated random graphs in the \emph{microcanonical} ensemble, with $N$
vertices and $L=N\alpha/2$ edges ($L$ is rounded to the nearest integer
value).  In the microcanonical ensemble the total number $L$ of edges is
fixed (in contrast to the \emph{canonical} ensemble in which $L$
fluctuates).

As we want to simulate graphs with $N$ up to $10^6$ and with an average
connectivity $\alpha$ of order $O(1)$, we must use an algorithm which
requires computer memory and time of order $O(N)$, not
$O(N^2)$.  With the microcanonical ensemble, the program is simpler: a
random graph is obtained by choosing at random $L$ distinct edges among all
the possible edges.  From a Monte-Carlo point of view, the microcanonical
ensemble has another advantage: the measurements fluctuate less.

As the graph (or equivalently its adjacency matrix) is very sparse, it is
stored in an array $T$ of $2L$ integers, indexed by an array $K$ of $N+1$
integers; the set of vertices adjacent to the vertex $v$ is the array
section $\{T(i)\}$ where $K(v) \leq i < K(v+1)$.  Then the connectivity of
$v$ is $K(v+1)-K(v)$.  This defines the array $K$, with the rules $K(1) =
1$ and $K(N+1) = 2L+1$.

Note that each edge $\{v,w\}$ appears twice in $T$: once for $v$ and once
for $w$.  This requires twice more memory than a storage method exploiting
the symmetry of the matrix (in which the edges appear only once), but the
computational task is faster because the adjacent vertices of a given
vertex are simply obtained from arrays $T$ and $K$.

The leaf removal process is done leaf by leaf.  Each time a leaf is
removed, the adjacent vertices are examined: if new leaves appear, there
are added to a list of potential leaves to be considered later.  Each
elementary leaf removal requires a computational time of order $O(1)$ and
not $O(N)$.  So the computational time for the global leaf removal is
proportional to the number of removed leaves, which is bounded by $N/2$.
Then the total computational time for one random graph (generation and leaf
removing) is $N$ times a function of $\alpha$.

For each random graph, we have measured the number of isolated points
$|I|$, the size (number of vertices) of the core $N_c$, the number of edges
of the core $L_c$ and consequently the average connectivity of the core
$2L_c/N_c$: there are estimators of $N i(\alpha)$, $N c(\alpha)$, $N
l(\alpha)$ and $\alpha_{eff}$, as defined before.  As usual in Monte-Carlo
simulations, averages have been done over many random graphs, and their
confidence intervals (or error bars) are estimated by the variance of the
measurements.

For each value of $\alpha$, we have generated $10\,000$ graphs of size
$N=100$, $N=1000$ and $N=10\,000$, and $1000$ graphs of size $N=100\,000$
and $N=1\,000\,000$. At the transition value $\alpha=e$, $10\,000$ graphs
have been generated for all the sizes.  The whole computation takes a few
days on a medium Sun workstation without special optimization.

\subsection{Monte-Carlo results}

\begin{figure}
  \centering
  \includegraphics[width=\figwidth, height=0.9\textheight, keepaspectratio]
                  {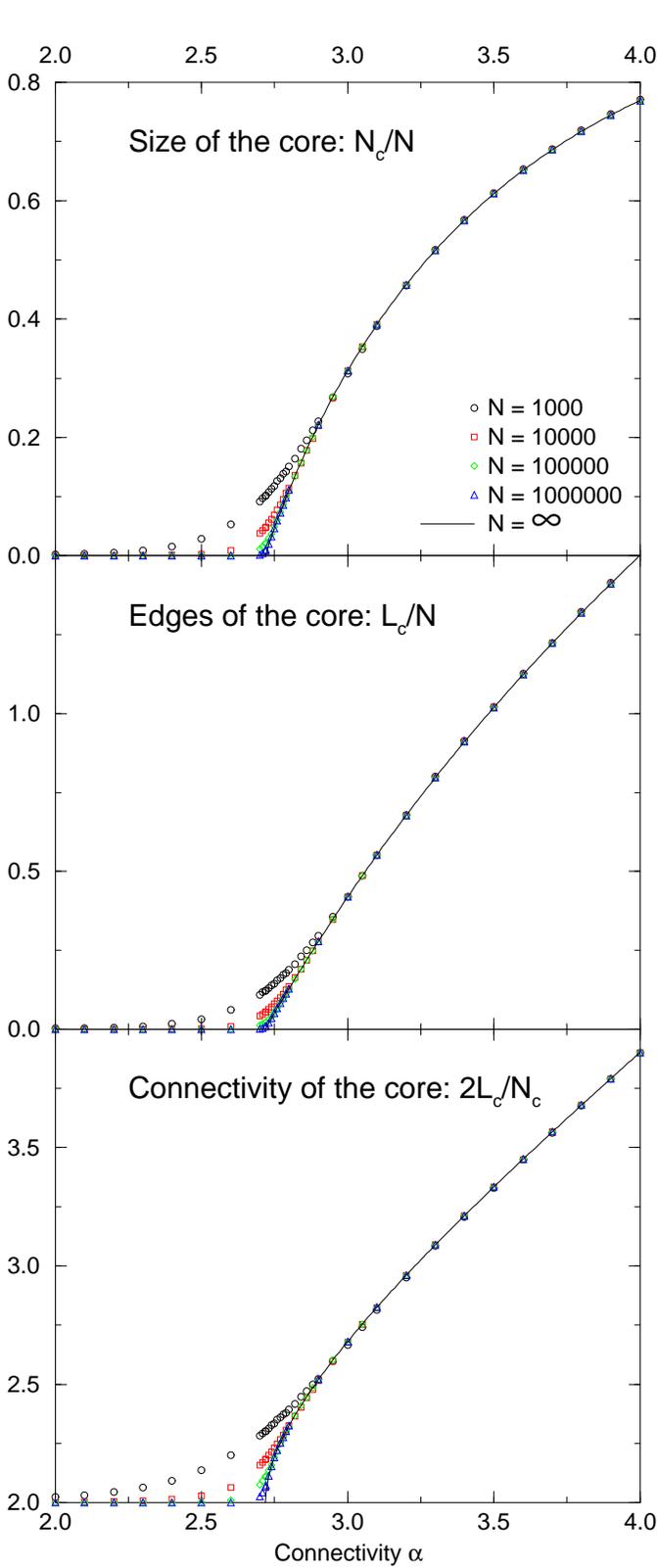}
  \caption{\em 
    Monte-Carlo averages (symbols) and analytical results (solid line) for
    the size, edges and average connectivity of the core.
  }
  \label{fig:moy}
\end{figure}

On Fig.~\ref{fig:moy}, Monte-Carlo averages of $N_c/N$, $L_c/N$ and
$2L_c/N_c$ are compared with the infinite $N$ results, $c(\alpha)$,
$l(\alpha)$ and $\alpha_{eff}$.  Errors bars are not drawn because they are
smaller than the size of symbols.  This figure is a typical case of a
second order transition.  Far from the transition, differences between
finite $N$ and thermodynamic (i.e. $N=\infty$) functions are small.  Finite
size effects are at least of order $1/N$, because the analytical
calculations do not take into account the loops of finite size: their
number is $O(1)$, so their
contributions are $O(1/N)$. The simplest example is the
``triangle'' subgraph made of three vertices and three edges, not
connected to the rest of the graph.  Obviously, the triangles have no leaf
and belong to the core: their contribution to $N_c$ is $(\alpha
e^{-\alpha})^3/2$.  

We have verified that finite size effects are of
order $O(1/N)$ (but not larger) for $c(\alpha)$ and $l(\alpha)$ far from
the transition.  For $\alpha_{eff}$, this is probably true but less clear
because fluctuations are stronger.
On the other hand, in the critical region $\alpha \sim e$, finite size
effects are larger and some critical exponents can be defined.  They are
discussed later.

\begin{figure}
  \centering
  \includegraphics[width=\figwidth, height=0.9\textheight, keepaspectratio]
                  {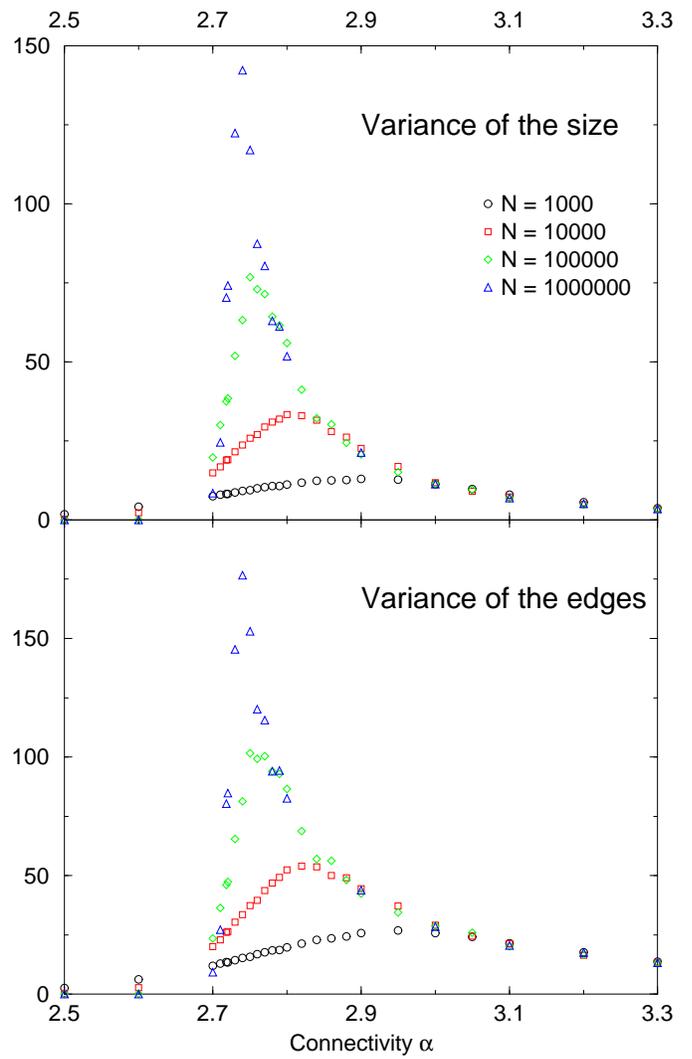}
  \caption{\em 
    Monte-Carlo estimations of the variance of the size of the core
    $\chi_c(\alpha)$ and the variance of the number of edges of the core
    $\chi_l(\alpha)$.
  }
  \label{fig:var}
\end{figure}

We have also examined the variances of the size, number of edges and
average connectivity of the core.  For $\alpha$ not too close to $e$ and
large $N$, we expect that the fluctuations (square root of
the variance) are of order $O(\sqrt{N})$ for $N_c$ and $L_c$, and
$O(1/\sqrt{N})$ for $\alpha_{eff}$.  So to obtain a large $N$ limit,
we define $\chi_c(\alpha) \equiv \Var(N_c)/N$, $\chi_l(\alpha) \equiv
\Var(L_c)/N$ and $\chi_\alpha(\alpha) \equiv N\ \Var(2L_c/N_c)$.  These
quantities are analogous in the spin models to the magnetic susceptibility
(equivalent to the fluctuations of the magnetization).

\begin{figure}
  \centering
  \includegraphics[width=\figwidth]{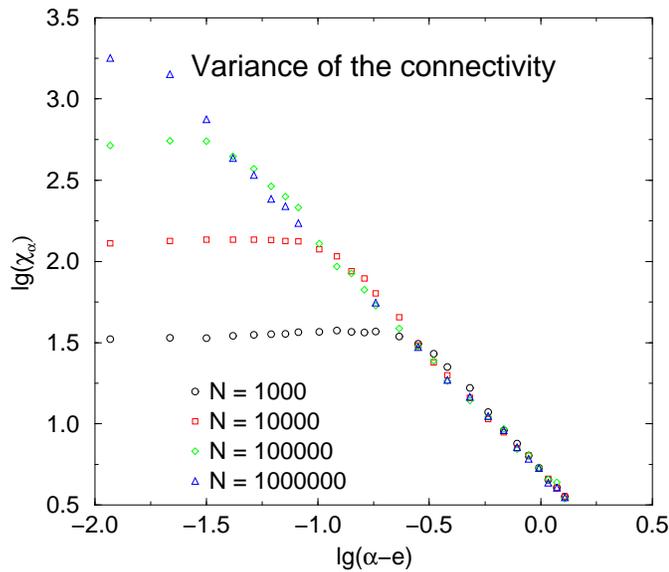}
  \caption{\em Monte-Carlo estimations of $\chi_\alpha(\alpha)$ (variance
  of the average connectivity of the core) versus $(\alpha-e)$.  The
  axes are labeled by decimal logarithms.}
  \label{fig:av}
\end{figure}

On Fig.~\ref{fig:var}, Monte-Carlo estimations of $\chi_c$ and $\chi_l$ show
that $\chi_c(\alpha)$ and $\chi_l(\alpha)$ have a finite limit for
$N=\infty$ when $\alpha>e$, a vanishing limit when $\alpha < e$ and diverge
when $\alpha$ approaches the critical value $e$.  
By analogy with $c(\alpha) \sim
l(\alpha) \sim (\alpha-e)$ and $\alpha_{eff}-2 \sim (\alpha -e)^{1/2}$ when
$\alpha\to e^{+}$, power laws are expected for the divergences.  So, we
define two critical exponents $\rho$ and $\rho'$ by
\begin{eqnarray}
  \chi_c(\alpha) \sim \chi_l(\alpha) & \sim & (\alpha-e)^{-\rho}, 
        \label{eq:rho} \\
  \chi_\alpha(\alpha)                & \sim & (\alpha-e)^{-\rho'},
        \label{eq:rho2}
\end{eqnarray}
when $\alpha\to e^{+}$.  The exponent $\rho$ could be numerically measured
by plotting $\log(\chi_c)$ or $\log(\chi_l)$ versus $\log(\alpha-e)$.
Unfortunately, this gives poor results because the curvature of the plot is
too important, with a slope $\rho$ changing from 1 to 0.5.  But it is
possible for $\alpha_{eff}$.  Fig.~\ref{fig:av} is a log-log plot of
$\chi_\alpha(\alpha)$ versus $(\alpha-e)$.  Symbols are lined up correctly
and the global slope gives the estimation
\[ 
  \rho' = 1.5(1).
\]

\begin{figure}
  \centering
  \includegraphics[width=\figwidth, height=0.9\textheight, keepaspectratio]
                  {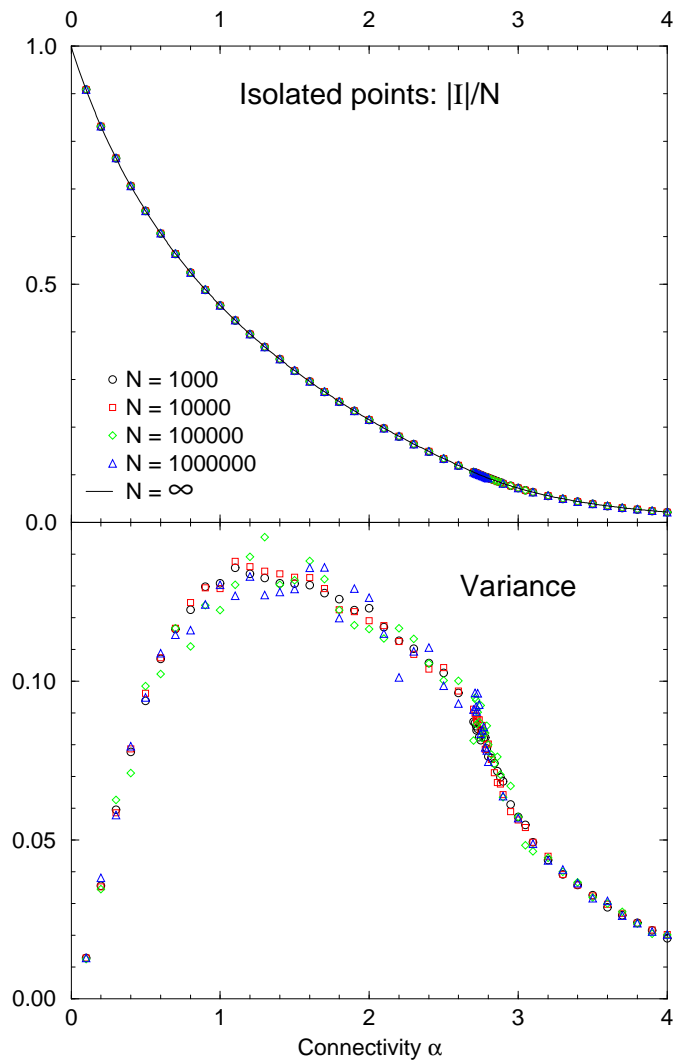}
  \caption{\em 
    Monte-Carlo averages $|I|/N$ and variance $\chi_i(\alpha)$ of the
    number of isolated points.  The solid line is the analytical result
    $i(\alpha)$ for $N = \infty$.
  }
  \label{fig:isole}
\end{figure}

The studies of isolated points are resumed on Fig.~\ref{fig:isole}.
Monte-Carlo averages of $|I|/N$ are compared with the infinite $N$ results,
$i(\alpha)$: errors bars and finite size effects are so small that they are
not visible.  On the other hand, the variance $\chi_i(\alpha) =
\Var(|I|)/N$ shows bigger statistical fluctuations, but the finite size
effects remain small.  This variance does not diverge anywhere.  However we
see a cusp when $\alpha \to e^+$ compatible with
\[
  \chi_i(e) - \chi_i(\alpha) \sim (\alpha - e)^\tau
\]
with estimations $\chi_i(e) = 0.095(5)$ and $\tau = 0.6(1)$. As $\tau < 1$,
the first derivative is infinite at $\alpha = e^+$.

\subsection{Finite size scaling}

We now concentrate on the finite $N$ behavior, first exactly at the
transition $\alpha = e$ and then in the critical region around this
transition.

\begin{figure}
  \centering
  \includegraphics[width=\figwidth]{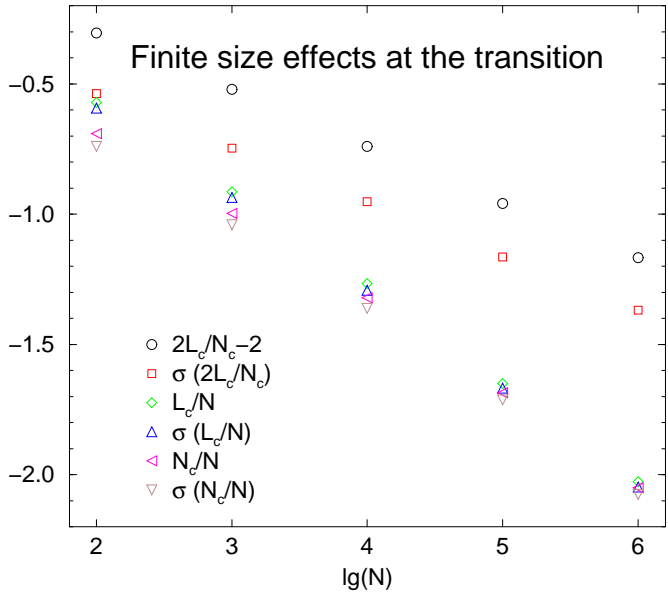}
  \caption{\em 
    Top to bottom: the average connectivity (mean and width), the number of
    edges (mean and width) and the size (mean and width) of the core versus
    the size $N$ of the random graph.  The axes are labeled by decimal
    logarithms.  The negative slopes are measurements of $-\phi$ (for
    the connectivity) and $\omega-1$ (for size and edges).
  }
  \label{fig:omega}
\end{figure}

By analogy with the classical percolation transition at $\alpha=1$ where
the size of the largest connected component is~\cite{erdos} of order
$O(N^{2/3})$ and its average connectivity is $2 + O(1/N^{2/3})$, we
postulate the existence of other critical exponents $\omega$ and $\phi$
defined by
\begin{eqnarray}
  N_c \sim L_c & \sim & N^\omega, \label{eq:omega}\\
  \alpha_{eff} - 2 & \sim & N^{-\phi} \label{eq:phi}
\end{eqnarray}
when $N \to \infty$ at $\alpha=e$.  This hypothesis is tested on
Fig.~\ref{fig:omega}: data are correctly fitted by power-laws with
\[
  \omega = 0.63(1) \mbox{\ \ and \ \ } \phi = 0.21(1). 
\]
Of course, if the large $N$ behavior is modified by a (power of a)
logarithmic function, the true values of the exponents are different than
their apparent values when $N$ is large but finite.  Here these exponents
are determined by considering the averages of the Monte-Carlo measurements.
The width $\sigma$ of their distributions are also plotted on
Fig.~\ref{fig:omega}: means and widths have similar slopes.  Consequently
\begin{equation}
  \chi_c(e) \sim \chi_l(e) \sim N^{2\omega-1}
  \mbox{\ \ and \ \ }
  \chi_\alpha(e) \sim N^{1-2\phi}
  \label{eq:chin}
\end{equation}
diverge when $N \to \infty$ at $\alpha=e$.  

\begin{figure}
  \centering
  \includegraphics[width=\figwidth, height=0.9\textheight, keepaspectratio]
                  {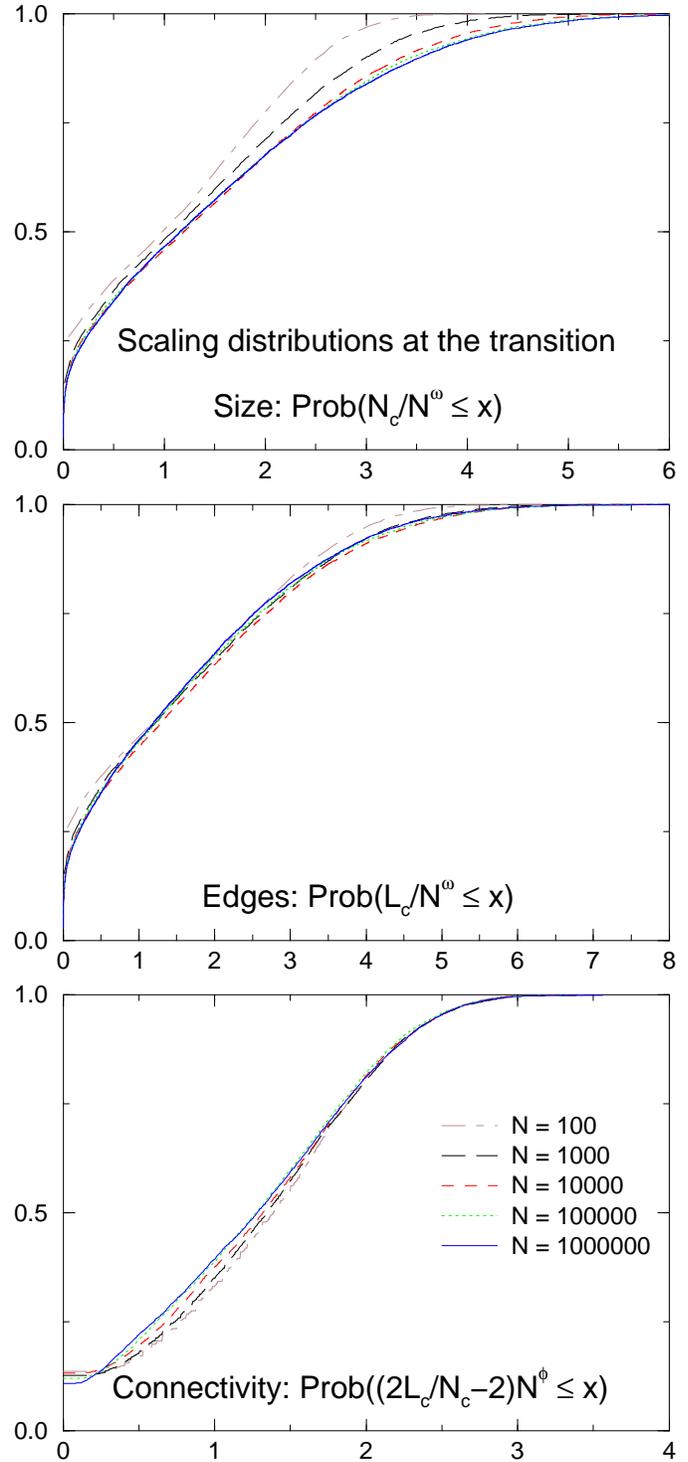}
  \caption{\em
    Cumulative distribution functions of the size, the number of edges and
    the average connectivity of the core, as functions of their respective
    scaled variables.  We have set $\omega = 0.63$ and $\phi = 0.21$.
  }
  \label{fig:cumul}
\end{figure}

As widths and means of the Monte-Carlo measurements are of the same order,
the distributions remain broad in the large $N$ limit at the transition.
On the contrary, when $\alpha \neq e$, the distributions are sharp.  On
Fig.~\ref{fig:cumul}, the \emph{cumulative} distribution functions
Prob($N_c/N^\omega \leq x)$, Prob($L_c/N^\omega \leq x)$ and
Prob$((2L_c/N_c - 2)N^\phi \leq x)$ are plotted as functions of the scaling
variable $x$ for $\alpha=e$.  We observe that the curves converge when $N$
is large to scaling distributions (independent of $N$) and this confirms the
hypotheses Eq.~(\ref{eq:omega}) and Eq.~(\ref{eq:phi}).

When $x$ becomes large, these scaling distribution functions decrease like
a Gaussian.  Consequently, they are not \emph{large} distributions in
the sense that every moment is finite, in agreement with
Eq.~(\ref{eq:chin}).  On the other side, these functions seem to be power
laws for small $x$.  This allows to define another exponent $\delta$ with
\begin{equation}
  \mbox{Prob}(N_c/N^\omega \leq x) \sim
  \mbox{Prob}(L_c/N^\omega \leq x) \sim  x^\delta
  \label{eq:delta}
\end{equation}
when $x \to 0$.  Our estimation is
\[
  \delta = 0.36(3).
\]
The numerical values suggest that $\omega + \delta = 1$, but we have
no argument to explain it.

By considering the probability that the core of a random graph is void
at $\alpha=e$, we measured a new exponent
\[
  \eta = 0.25(1)
\]
where $\eta$ is defined by
\[
  \mbox{Prob}(N_c=0) \sim N^{-\eta}.
\]
The limit $x\to 0$ in Eq.~(\ref{eq:delta}) gives the conjectured relation
$\eta = \omega \delta$, which is numerically acceptable.  With the
hypothesis $\omega + \delta = 1$, it gives $\eta = \omega (1-\omega)$.

We have also considered Prob$(L_c = N_c)$, i.e. the probability that the
average connectivity of the core is exactly 2.  In this case, the core is
made of one or several simple loops without branching.  The Monte-Carlo
study indicates that the large $N$ limit could be a pure number: 0.12(2).
More intensive simulations are needed to confirm (or invalidate) this
result.

More relations between critical exponents can be obtained by using the
finite size scaling hypothesis~\cite{barber}: in the vicinity of the
transition, the behavior of finite random graphs is
determined by the scaling variable
\[
  y = (\alpha-e) N^\theta
\]
where $\theta$ is a positive  \emph{scaling exponent}. 

First we shortly resume the scaling theory for a general quantity
$Q(N,\alpha)$, for size $N$ and connectivity $\alpha$.  For $N=\infty$, let
us suppose that
\[
  Q(\alpha) \sim (\alpha-e)^\gamma
\]
when $\alpha\to e^{+}$ ($\gamma$ could be positive or negative).  Then we
expect in the critical region that
\[
  Q(N,\alpha) \sim N^{-\gamma\theta}\ \tilde{Q}(y)
\]
where the scaling function $\tilde{Q}(y)$ is defined by
\[
  \tilde{Q}(y) \equiv 
  \lim_{N\to\infty}\ N^{\gamma\theta}\ Q(N,\ e + y/N^\theta),
\]
which behaves as
\[
  \tilde{Q}(y) \stackrel{y \to +\infty}{\sim} y^\gamma.
\]
As $y=0$ exactly at the transition $\alpha = e$, 
\[
  Q(N,e) \sim N^{-\gamma\theta}.
\]
So the exponent of finite size effects at the transition and the exponent
of critical behavior for $N=\infty$ in the vicinity of the transition are
related by $\theta$.  This remark is useful only if different quantities
share the same $\theta$.  For usual models of statistical physics with a
2-D or 3-D geometry (like classical spin systems), the exponent $\theta$
describes the divergence of the correlation length $\xi$.  So the
uniqueness of $\xi$ implies the uniqueness of $\theta$.  Unfortunately for
random graphs, we have no equivalent length and no simple phenomenological
interpretation for $\theta$.  However we shall assume that $\theta$ is
unique.

As we have computed exact $N=\infty$ formul{\ae} in
Sect.~\ref{sec:coreperco}, we can directly study $Q(N,\alpha) - Q(\alpha)$,
i.e. the finite size effects.  The scaling function is now $\tilde{Q}(y) -
y^\gamma$ and is maximal around $y=0$.  From a numerical point of view, the
analysis becomes easier than the one of the monotonic function
$\tilde{Q}(y)$.

\begin{figure}
  \centering
  \includegraphics[width=\figwidth]{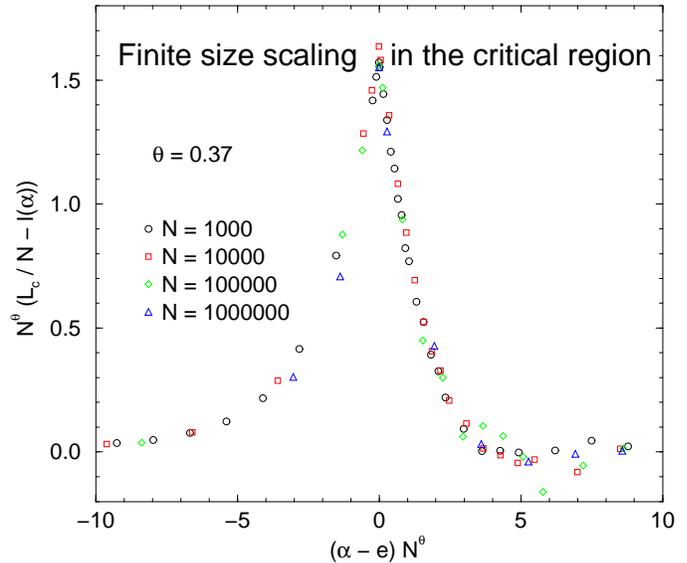}
  \caption{\em 
    Finite size scaling for the edges of the core in the critical region.
  }
  \label{fig:fss}
\end{figure}

Let us now consider $N_c/N$ and $L_c/N$.  As $c(\alpha) \sim l(\alpha) \sim
(\alpha-e)$ when $\alpha \to e^+$, for these quantities $\gamma=1$.  Then
\[ 
  \theta = 1 - \omega. 
\]
On Fig.~\ref{fig:fss}, with the choice $\theta = 0.37$ (induced by the
numerical measure of $\omega$), $N^\theta (L_c/N - l(\alpha))$ is plotted
versus $y= (\alpha-e) N^\theta$.  We see that data are well superposed:
they draw the scaling function.  Note that $\theta$ is the unique fitting
parameter for this figure.

Let us now consider the variances $\chi_c(\alpha)$ and $\chi_l(\alpha)$.
Using Eq.~(\ref{eq:rho}) and Eq.~(\ref{eq:chin}), the finite size scaling
hypothesis gives the new relation
\[
  \rho \theta = 2\omega - 1.
\]

The same analysis with the average connectivity of the core can be done.
As $\alpha_{eff} -2 \sim (\alpha-e)^{1/2}$, the corresponding $\gamma =
1/2$.  Using Eq.~(\ref{eq:phi}), the scaling relation is
\[
  \theta = 2 \phi.
\]
For the variance of the connectivity $\chi_\alpha$, Eq.~(\ref{eq:rho2}) and
Eq.~(\ref{eq:chin}) give 
\[
  \rho' \theta = 1 - 2 \phi.
\]
By eliminating $\theta$, other relations are obtained: $\rho' = \rho+1$ and
$2\phi+\omega=1$.

\begin{table}
  \centering \tabcolsep 4.5pt
  \begin{tabular}{|c|c|cc|cc|c|}
    \hline
    & $\theta,\ \delta$ & $\omega$ & $\phi$ & $\rho$ & $\rho$' & $\eta$ \\
    \hline
    MC & 0.36(3) & 0.63(1) & 0.21(1) &        & 1.5(1) & 0.25(1)\\
    \hline
    a & 1/3     & 2/3     & 1/6     & 1      & 2      & 2/9    \\
    b & 0.37    & 0.63    & 0.185   & 0.70   & 1.70   & 0.233  \\
    c & 2/5     & 3/5     & 1/5     & 1/2    & 3/2    & 24/100 \\
    d & 0.42    & 0.58    & 0.21    & 0.38   & 1.38   & 0.244  \\
    \hline
  \end{tabular}
  \caption{\em
    Critical exponents for the geometric phase transition when the average
    connectivity of a large random graph is $\alpha=e$. The line ``MC''
    displays the results of Monte-Carlo simulations.  The lines ``a--d''
    are a few sets of values compatible with scaling relations.  The line 
    ``c'' has our preference (see text).
  }
  \label{tab:exponents}
\end{table}

With these four scaling relations and the Monte-Carlo determinations, we
will now try to conjecture the \emph{exact} values of these exponents.
Table~\ref{tab:exponents} resumes the following considerations.  The
results of Monte-Carlo simulations are given in line ``MC''.  Other lines
are suggestions for sets of exponents compatible with the four scaling
relations.

The line ``b'' is obtained by using the numerical determination of $\omega$
and the scaling relations.  In particular, it gives $\theta = 0.37(1)$.
The line ``d'' uses the numerical determination of $\phi$; it gives $\theta
= 0.42(2)$.  As the difference between these two values of $\theta$ is
about twice larger than the uncertainty, we cannot definitely conclude
whether the size and the connectivity of the core share the same
scaling exponent $\theta$ or not.

The line ``a'' is obtained by assuming that $\omega=2/3$ and $\theta =
1/3$, which are the values~\cite{erdos} of the corresponding exponents
for the classical percolation of random graphs at $\alpha=1$.  This
hypothesis seems incompatible with the Monte-Carlo estimations of
$\phi$ and $\rho'$.  Furthermore the average connectivity of the giant
component near the classical percolation transition behaves with $2 +
O((\alpha-1)^2)$ --- to be compared with $2 +O((\alpha-e)^{1/2})$ for
the core --- and consequently the corresponding exponent $\phi$ is
$2\theta = 2/3$ (but not $\theta/2 = 1/6$).  

 This gives strong evidence that the analogy between exponents of
percolation and core transitions cannot be complete and that the
effective low energy field theory descriptions in the vicinity of the
transition are different. In particular, they may well have a different
upper critical dimension. 

The line ``c'' assumes that the exponent $\rho'= 1.5(1)$ is exactly $3/2$.
This is very attractive because exponents are simple rational
fractions and the value $\theta = 2/5$ is between the numerical estimations
$0.42$ and $0.37$.

Of course, nothing in the theory of critical phenomena requires that
critical exponent are rational numbers with small numerators (for a recent
example, see Ref~\cite{dgg}).  However, if we want conciliate
numerical simulations, theoretical considerations and simple rational
fractions, we are led to conjecture $\omega = 3/5$, $\phi=1/5$, 
$\rho=1/2$, $\rho'=3/2$, $\delta=\theta=2/5$ and $\eta=6/25$.

To reduce the uncertainties in Monte-Carlo simulations, bigger size $N$ are
needed, in particular in the case where the large $N$ behavior would be
affected by logarithmic laws. Moreover, we hope to progress in analytical
methods for calculating these exponents as well.

\section{Applications}
\label{sec:applis}

We now discuss two applications of   the structure of
the core:   the first one to the conductor-insulator
transitions in random graphs and the second one to combinatorial
optimization problems.

\subsection{Localization on random graphs}
\label{sec:loc}

We denote by $Z(G)$ the dimension of the kernel (the subspace of
eigenvectors with eigenvalue $0$) of the adjacency matrix of the graph
$G$. It is known~\cite{cvetkovic} that $Z(G)$ is invariant under leaf
removal (see Ref.~\cite{baugolkerntree} for a proof and an application to
random trees).  As the adjacency matrix is block-diagonal with one block
per connected component, $Z(G)$ is additive on connected components. These
two properties imply that
\[ 
  Z(G)=Z(C)+Z(I)=Z(C)+|I| \leq N_c+|I|.
\]
The last equality is because the adjacency matrix vanishes for a
collection of isolated points. This analysis applies to any graph, and
remains valid after averaging. Even if the probability distribution is
not that of a random graph, we see that as soon as leaves appear with
a non vanishing weight (this is true for instance if the probability
distribution is that of a lattice with impurities), the spectrum of
the adjacency matrix has a delta peak at the origin. However, the
fact that, as we show below for random graphs, leaf removal accounts
for the full weight of this delta peak seems rather non generic.

Taking the average of these formul\ae\ for random graphs and using our
results on the core, we get that $Z(G) = Nz(\alpha)+o(N)$ for a large
random graph $G$ with average connectivity $\alpha$, with
\begin{eqnarray}
  z(\alpha) = i(\alpha) \quad \mathrm{for} \quad \alpha \leq e ,
  \label{ineqzi} \\
  i(\alpha) \leq z(\alpha) \leq c(\alpha) +i(\alpha) \quad \mathrm{for}
  \quad \alpha > e .  \nonumber
\end{eqnarray}

It has been argued in Ref.~\cite{baugolloctrans} that
\begin{equation}
  \label{zegal}
  z(\alpha)=\frac{A+B+AB}{\alpha}-1  
\end{equation}
for all values of $\alpha$. Combined with our present results, this means
that
\[
  z(\alpha)=i(\alpha) \quad \mathrm{for \; all \; values \; of} \
  \alpha.
\]
We may interpret Eq.~(\ref{ineqzi}) as an independent proof of
Eq.~(\ref{zegal}) for $\alpha \leq e$ and we may also infer that the
adjacency matrix of the core of a random graph at $\alpha >e$ has a kernel
of size $o(N)$.

In Ref.~\cite{baugolloctrans}, it was shown that $e$ is in a domain of the
$\alpha$ parameter for which delocalized vectors are responsible for a
finite fraction of the size of the kernel. Imagine that we start to
increase $\alpha$ very slowly from $\alpha=e$ by adding randomly new edges
one by one to the random graph. We watch the competition between the core
(which, we have seen, carries few elements in the kernel), the localized
eigenvectors in the kernel and the delocalized ones.  The competition
between the core and the full kernel is not very strong: when $n$ edges,
with $1 << n << N$, are added to the graph, the core grows in average of
$24\ e^{-2}n$ vertices, while $e^{-2}n$ vectors in the kernel are
lost. However, by the results of Ref.~\cite{baugolloctrans}, about $ n^2$
delocalized eigenvectors disappear and $ n^2$ localized eigenvectors
replace these. It is intuitive that delocalized eigenvectors in the kernel,
which live on large structures on the random graph, have a high probability
to be perturbed by the growing core. But the precise mechanism by which
their extinction is almost compensated by new localized vectors in the
kernel remains to be elucidated.

The concept of leaf removal process can also be used to analyze the
localization-delocalization transitions that occur at $\alpha _d\approx
1.42153$ and $\alpha_r\approx 3.15499$.  As shown in
Ref.~\cite{baugolloctrans}, the localized eigenvectors in the kernel live
on definite structures that can be drawn on the random graph.  These
structures are finite (connected) trees that
\begin{itemize}

\item can be bicolored brown-green in such a way that all vertices with 0
or 1 neighbor are green and all neighbors of the green vertices in the
random graph belong to the tree; the neighbors of the brown vertices on the
other hand, can be anywhere on the random graph.

\item are maximal, i.e. are not part of a larger tree with the same
properties. Observe that each isolated point is maximal.

\end{itemize}
We can put marks on all vertices belonging to such structures and build
histories of leaf removal processes such that the initial steps remove only
marked vertices, and after these steps, the only remaining marked vertices
are now isolated.

Then if $\alpha$ is small ($\alpha \leq \alpha _d$) or large ($\alpha \geq
\alpha_r$), the number of isolated marked points is $Ni(\alpha)+o(N)$. This
implies in particular that at most $o(N)$ bunches of  the remaining
graph contain more than one leaf.

On the other hand, if $\alpha \in ]\alpha _d,\alpha _r[$ the number of
isolated marked points is less than $Ni(\alpha)$: a number of order $N$ of
non-trivial bunches will have to appear somewhere during the rest of the
leaf removal process.

\subsection{Vertex covers and matchings}
\label{sec:guard}

Apart from the size of the kernel, several other interesting quantities
attached to graphs behave rather simply under leaf removal. We mention two,
which are related to combinatorial optimization problems.

\textbf{Vertex cover}: A vertex cover of a graph is a subset of the
vertices containing at least one extremity of every edge of the graph. We
denote by $X(G)$ the minimal size of a vertex cover of a graph $G$.

 There is a nice ``practical'' interpretation of $X(G)$.  Imagine that the
edges of the graph are the (linear) corridors of a museum, the vertices
corresponding to ends of corridors. A guard sitting at a vertex can control
all the incident corridors. So $X(G)$ is the minimum number of guards
needed to control all corridors of the museum. 

\textbf{Edge disjoint subset, matching}: An edge disjoint subset is a
subset of the edges such that no two edges in the subset have a vertex
in common.  This is also called a matching. We denote by $Y(G)$ the
maximal size of edge disjoint subset in a graph $G$. Finding such a
maximal edge disjoint subset is called the matching problem. 

The determination of $X(G)$ or $Y(G)$ for a given $G$ are archetypal of two
classes of optimization problems. While it is known that the matching
problem can be solved in polynomial time (see e.g.  Ref.\cite{pitteletal}
and references therein), the museum guard problem is in the
Non-deterministic Polynomial (NP) class because no polynomial time
algorithm is known to solve it (and such an algorithm is not expected to
exist, this is related to the famous P$\neq$NP conjecture), but given a
candidate solution, it is easy to check in polynomial time that it is
correct.

When $G$ is a large random graph, one may ask for thermodynamic
solutions of these problems, when only the extensive contributions to 
$X(G)$ or $Y(G)$ are considered as relevant. This leads to the
following definition: 

 \textbf{Vertex cover fraction $x(\alpha)$, matching
  fraction $y(\alpha)$}: For fixed $\alpha$, the vertex cover fraction 
$x(\alpha)$ and the matching fraction $y(\alpha)$ are the limits of
the averages of $X(G)/N$ and $Y(G)/N$ when $G$ is a random graph of
size $N$ and $N \rightarrow \infty$. 

Let us note however that the one can probably exhibit combinatorial
optimization problems for which the thermodynamic solution can be
obtained in polynomial time, but taking into account the non-extensive
remainder is prohibitively long.

The replica trick has been used by Hartmann and Weigt to obtain a
lower bound for $x(\alpha)$ in a series of
papers~\cite{hartweig1,hartweig2,hartweig3}. They have shown that for
$\alpha \leq e$, the replica symmetric solution is stable, whereas it
become unstable for $\alpha >e$. The replica symmetric solution leads
to
\[ 
  x(\alpha)=1-\frac{2W+W^2}{2\alpha} \quad \mathrm{for} \quad \alpha \leq e.
\]
This relation has to break down somewhere, because a result of Frieze
\cite{frieze} implies that 
\[
  x(\alpha)=1-\frac{2}{\alpha}(\log \alpha- \log \log \alpha -\log 2
  +1)+o(\frac{1}{\alpha}), 
\]
whereas $W\sim \log \alpha$ for large $\alpha$, so that the asymptotics of
$1-\frac{2W+W^2}{2\alpha}$ starts with $1-\frac{\log ^2\,
\alpha}{2\alpha}$.  Weigt and Hartmann~\cite{hartweig3} have also used a
good algorithm to get an approximation of a minimal vertex cover. The idea
is essentially to look for a vertex with a maximal number of incident
corridors and put a guard there. Then remove the site and the adjacent
corridors and iterate. This is always fast, but gives only an upper bound
for $X(G)$. This can be refined, but then the algorithm needs a very long
time when $\alpha >e$.

\vspace{.3cm} 

We show how leaf removal can be applied to the museum guard problem.  If
$(v,w)$ is a leaf of $G$, there is a minimal vertex cover with a guard at
$v$. This is because any vertex cover has a guard at $v$ or at $w$, and a
guard at $v$ makes the guard at $w$ useless. So if a minimal vertex cover
has a guard at $w$, moving it to $v$ yields another minimal vertex
cover. Isolated vertices do not need guards. The leaf removal of $(v,w)$
leading from $G$ to $G'$ removes exactly the corridors controlled by the
guard at $v$. Hence $X(G)=X(G')+1$.

An analogous argument applies to maximum edge disjoint subsets.  Indeed, if
$(v,w)$ is a leaf of $G$, there is a maximal edge disjoint subset that
contains $\{v,w\}$. This is because if no edge of an edge disjoint subset
touches $v$, this edge disjoint subset is not maximal (it can be completed
with $\{v,w\}$), and if an edge disjoint subset has an edge that touches
$v$, replacing this edge by $\{v,w\}$ yields an edge disjoint subset of the
same size. The leaf removal of $(v,w)$ leading from $G$ to $G'$ removes,
apart from $\{v,w\}$, exactly the edges that cannot belong to any edge
disjoint subset containing $\{v,w\}$. Hence $Y(G)=Y(G')+1$.

Some general inequalities can be proved. For instance $X(G) \geq Y(G)$ (the
triangle is an example when the inequality is strict) and $Z(G)\geq
N-2X(G)$. However, $Z(G)-N+2Y(G)$ can have any sign (negative for the
triangle but positive for the square).

Anyway, at each leaf removal, two vertices are removed, so $Z(G)$,
$N(G)-2X(G)$ and $N(G)-2Y(G)$ are invariant under leaf removal. Moreover,
$X$ and $Y$ vanish for unions of isolated vertices.  To summarize,
\begin{eqnarray*} 
  X(G) & = & X(C)+\frac{N-N_c-|I|}{2} \\
  Y(G) & = & Y(C)+\frac{N-N_c-|I|}{2}. \\
  Z(G) & = & Z(C)+|I|. 
\end{eqnarray*}

Karp and Sipser \cite{karpsipser,pitteletal} have devised an approximate
algorithm to get a good (if not optimal) matching. There are two possible
transformations:

(1) Remove a leaf,

(2) Choose an edge at random, remove it together with its
extremities and all edges touching the extremities. 

\noindent and they are are performed according to the following rule:

At each step, until the graph is empty, do (1) if possible and if not,
do (2). So starting from $G$, one applies (1) until the core of $G$ is
obtained.  Then (2) is applied as long as no new leaf appears. As soon
as a graph with leaves appears, apply (1) to reach the core of this
new graph, and so on.

At each step an edge is singled out, and by construction, the set of all
these edges defines a matching, i.e. an edge disjoint subset.

When $G$ is a random graph with $\alpha <e$, the core is small ($o(N)$ for
large $N$). Thus,
\[ 
  x(\alpha)=y(\alpha)=\frac{1-z(\alpha)}{2}=1-\frac{2W+W^2}{2\alpha}
  \, \mathrm{for} \, \alpha \leq e,
\]
which gives in particular an independent proof of the result of Weigt and
Hartmann~\cite{hartweig1}.  Note that in this case, the approximate
algorithm is to put a guard at a vertex touching as many edges as possible,
then remove it and iterate, whereas the exact algorithm is almost the
opposite, namely, put a guard at the connected end of a leaf, remove the
leaf and iterate. Leaf removal gives a very fast algorithm (linear in $N$
if the graph is properly encoded) to construct a minimal vertex cover when
$\alpha <e$.

Karp and Sipser have shown that for a large random graph with
$\alpha > e$, their algorithm for matching finds with high probability
a matching of about $N_c/2$ edges in the core. This is a lower bound
for the matching number of the core, but at the same time, this is the
maximum possible. So this shows at once  that the core has a
thermodynamically perfect matching, and that their
algorithm is thermodynamically optimal. Hence
\[ y(\alpha)=1-\frac{A+B+AB}{2\alpha} \; \mathrm{for \; any} \, \alpha
\]  
so that the relation $y(\alpha)=\frac{1-z(\alpha)}{2}$ is valid for
every $\alpha$: the fact that the core does not contribute
thermodynamically to zero eigenvalues and the fact that it has a
thermodynamically perfect matching are closely related.

If $\alpha >e$, leaf removal stops while an extensive number of edges are
still present: this gives a lower bound for $x(\alpha)$ which is very poor
at large $\alpha$. But it seems clear that the replica symmetry
breaking~\cite{hartweig3} at $\alpha=e$ is tightly connected to the fact
that the structure of the core of a random graph is more complicated than
the structure of the parts eliminated by leaf removal, so that a more
refined Parisi order parameter is needed to describe the phase $\alpha >
e$.  While we have seen that the matching parameter $y(\alpha)$ and the
kernel-size parameter $z(\alpha)$ are well understood and closely related,
the exact evaluation of vertex cover parameter $x(\alpha)$ remains as a
challenge.

\section{Conclusions}

In this paper we have presented a physicist's analysis of a deep feature of
random graphs : a geometric second order phase transition with the
emergence of a core at threshold $\alpha =e$. This core is the residue when
leaves (i.e. points with a single neighbor and this neighbor) and isolated
points are iteratively removed. We have argued that the core is dominated
by a giant component.  The dominant contribution of the large $N$ behavior
of the relevant thermodynamic quantities was computed exactly by a direct
counting method.  We have studied numerically the finite size behavior of
the core, defined a variety of new critical exponents and obtained
approximations consistent with their mutual relationships. Our analysis
excludes the exponents of standard percolation in random graphs. Finally,
we have applied our results to the localization transition and to
combinatorial optimization problems on random graphs.

However, some more analytical or numerical work is needed to identify
without any doubt the exponents for the phase transition at $\alpha=e$.  An
open question is the interaction between the emergence of the core and the
delocalized eigenvectors of the adjacent matrix with eigenvalue 0.  A fine
study of the distribution of the sizes of the connected components of the
core could be done with Monte-Carlo simulations: for $\alpha > e$, we
expect a giant component, plus a finite number of finite components.
Moreover we have shown that the number of eigenvectors with eigenvalue 0
living on the core is $o(N)$ and numerical simulations could give the
precise asymptotics.

Finally, as the core percolation appears in a simple model of random
graphs, which is governed only by one parameter, the average connectivity
$\alpha$, we expect that this transition is \emph{universal} in the sense
that some characteristics of this transition (second order, critical
exponents, etc., but not the precise value of $\alpha = e$ at the
transition) could be seen in other models or real materials.

\vspace{.3cm}

Acknowledgments: We are very indebted to Graham Brightwell and Boris
Pittel for drawing our attention to the mathematical literature on
matching in random graphs and its relevance for our work.

\appendix
\section{\hspace{-.8cm} ppendix: The core is well-defined}

In this appendix, we show  by induction on the number $N$ of vertices
of $G$ that the property 
\begin{quote}
  ${\cal P}_N \equiv$ ``the number of isolated points $|I|$ after leaf
  removal and the core $C$ of a graph $G$ on $N$ vertices do not depend on
  the history''
\end{quote}
holds for any $N \geq 0$.

To start the induction, if $G$ has 0 or 1 vertex, there is no leaf hence
there is only one history, so ${\cal P}_0$ and ${\cal P}_1$ are true.
Suppose now that ${\cal P}_0,\cdots,{\cal P}_{N-1}$ are proved and take a
graph $G$ on $N\geq 2$ vertices. We distinguish several cases:

\begin{description}

\item[1] If $G$ has no leaf, there is only one history so ${\cal P}_N$ is
true for $G$.

\item[2] If $G$ has exactly one leaf, all histories start with the same
first leaf removal, lead to the same $G'$ for which ${\cal P}_{N-2}$ is
true by the induction hypothesis, so ${\cal P}_N$ is true for $G$.

\item[3] If $G$ has at least two leaves, we compare two histories:
\begin{eqnarray*}
  {\cal H}_1 & = & G,(v_1,w_1),G'_1,\cdots \quad \mathrm{and} \\
  {\cal H}_2 & = & G,(v_2,w_2),G'_2,\cdots.
\end{eqnarray*}

\item[3a] If $\{v_1,w_1\}=\{v_2,w_2\}$, $G'_1=G'_2$ to which the induction
hypothesis ${\cal P}_{N-2}$ applies, so that $C_1=C_2$ and $|I_1|=|I_2|$.

\item[3b] If $v_1=v_2$ but $w_1\neq w_2$ (the two leaves are distinct but
belong to the same bunch, this can happen only if $N \geq 3$), then $G'_1$
has $w_2$ as an isolated point, $G'_2$ has $w_1$ as an isolated point, but
$G'_1/\{w_2\}=G'_2/\{w_1\}=G''$, say. Further leaf removals can take place
only on $G''$, to which the induction hypothesis ${\cal P}_{N-3}$ applies,
so again, $C_1=C_2$ and $|I_1|=|I_2|$.

\item[3c] Suppose that $(v_1,w_1)$ and $(v_2,w_2)$ do not belong to the
same bunch. This can happen only if $N \geq 4$.  Then $(v_1,w_1)$ is a leaf
of $G'_2$ and $(v_2,w_2)$ is a leaf of $G'_1$. The graph obtained from
$G'_2$ by leaf removal of $(v_1,w_1)$ and the graph obtained from $G'_1$ by
leaf removal of $(v_2,w_2)$ are the same, because they are both equal to
$G''$, the subgraph of $G$ induced by $V/\{v_1,w_1,v_2,w_2\}$.  Take a
history ${\cal H}''$ for $G''$. It can be completed to give two histories
for G, ${\cal H}''_1=G,(v_1,w_1),G'_1,(v_2,w_2),{\cal H}''$ and ${\cal
H}''_2=G,(v_2,w_2),G'_2,(v_1,w_1),{\cal H}''$. The induction hypothesis
${\cal P}_{N-2}$ applies to $G'_1$ so ${\cal H}''_1$ and ${\cal H}_1$ have
to end with the same core and the same number of isolated points. The same
is true for ${\cal H}''_2$ and ${\cal H}_2$ because the induction
hypothesis ${\cal P}_{N-2}$ applies to $G'_2$, and also for ${\cal H}''_1$
and ${\cal H}''_2$ because the induction hypothesis ${\cal P}_{N-4}$
applies to $G''$. By transitivity, ${\cal H}_1$ and ${\cal H}_2$ end with
the same core and the same number of isolated points: $C_1=C_2$ and
$|I_1|=|I_2|$.

\end{description}
All possibilities have been examined, hence whatever the number of leaves
of $G$, ${\cal P}_N$ is true for $G$. This completes the induction
step.


\end{document}